\documentclass[superscriptaddress,showpacs,aps,prl,twocolumn]{revtex4}

\usepackage[dvips]{graphicx}
\usepackage{amsmath}
\begin{document}


\title{Noise enhancement due to quantum coherence in coupled quantum dots}

\author{G.~Kie{\ss}lich}

\affiliation{School of Physics and Astronomy,
University of Nottingham, Nottingham NG7 2RD, United Kingdom}
\affiliation{Institut f{\"u}r Theoretische Physik, Technische Universit{\"a}t
  Berlin, D-10623 Berlin, Germany}

\author{E.~Sch{\"o}ll}
\affiliation{Institut f{\"u}r Theoretische Physik, Technische Universit{\"a}t  
Berlin, D-10623 Berlin, Germany}

\author{T.~Brandes}
\affiliation{Institut f{\"u}r Theoretische Physik, Technische Universit{\"a}t
  Berlin, D-10623 Berlin, Germany}

\author{F.~Hohls}
\affiliation{Institut f{\"u}r Festk{\"o}rperphysik, Leibniz Universit{\"a}t Hannover,
  D-30167 Hannover, Germany}

\author{R.~J.~Haug}
\affiliation{Institut f{\"u}r Festk{\"o}rperphysik, Leibniz Universit{\"a}t Hannover,
  D-30167 Hannover, Germany}

\begin{abstract}
We show that the intriguing observation of noise enhancement in the charge transport 
through two vertically coupled quantum dots can be explained by the interplay of 
quantum coherence and strong Coulomb blockade. We demonstrate that this novel mechanism 
for super-Poissonian charge 
transfer is very sensitive to decoherence caused by electron-phonon 
scattering as inferred from the measured temperature dependence.
\end{abstract}

\maketitle
The direct manifestation of quantum coherence in time-averaged transport observables of 
nanoscale conductors such as current or low-frequency current fluctuations is of great 
benefit since it allows for the study of decoherence in a straightforward manner.
Quantum dots (QDs) coupled in series provide a perfect system to explore such effects since the quantum 
superposition of states in different dots has an immediate influence on the charge transfer \cite{WIE03}.   
The effect of quantum coherence has been experimentally addressed mostly with respect to the average 
or time-dependent currents (e.g in Ref.~\cite{HAY03}), whereas until now current fluctuations as a 
very sensitive diagnostic tool \cite{BLA00,AGU04,FUJ06a} have not been  investigated for this purpose.

In this Letter we demonstrate that our recent low-frequency shot noise data provide such a tool and 
are in fact a direct indicator of quantum coherent coupling between two layers of self-assembled InAs QDs 
\cite{BAR06}. In our device, QD stacks are formed such that vertical tunneling through three barriers 
becomes possible. The corresponding current vs. applied bias voltage exhibits sharp peaks caused by the 
alignment of levels in a single QD stack. At the edges of these resonances the noise is enhanced above the 
Poissonian value for uncorrelated electron transfer $2e\langle I\rangle$ ($e$ is the unit charge and 
$\langle I\rangle$ is the average current). 

We find a striking agreement between our data  and our calculations of this novel mechanism 
based on a quantum master equation for super-Poissonian charge transport
in a single QD stack, which shows that contrary to the common view
capacitive coupling between the stacks is not needed to explain the observed
features. Typically, such noise behavior is 
associated with electron bunching in the charge transfer and  has therefore  received remarkable interest in 
recent experimental \cite{IAN98,KUZ98,SAF03,GUS06a,ZAR07} and theoretical work
\cite{KIE03b,THI04,COT04,BEL05,DJU05,AGH06a}. 
We also show that the decoherence due to electron-phonon scattering causes a temperature dependence 
that agrees with the measurements, allowing us to clearly identify the crossover between coherent and 
sequential tunneling.

\begin{figure}[b]
  \begin{center}
\raisebox{2mm}{
    \includegraphics[width=0.12\textwidth]{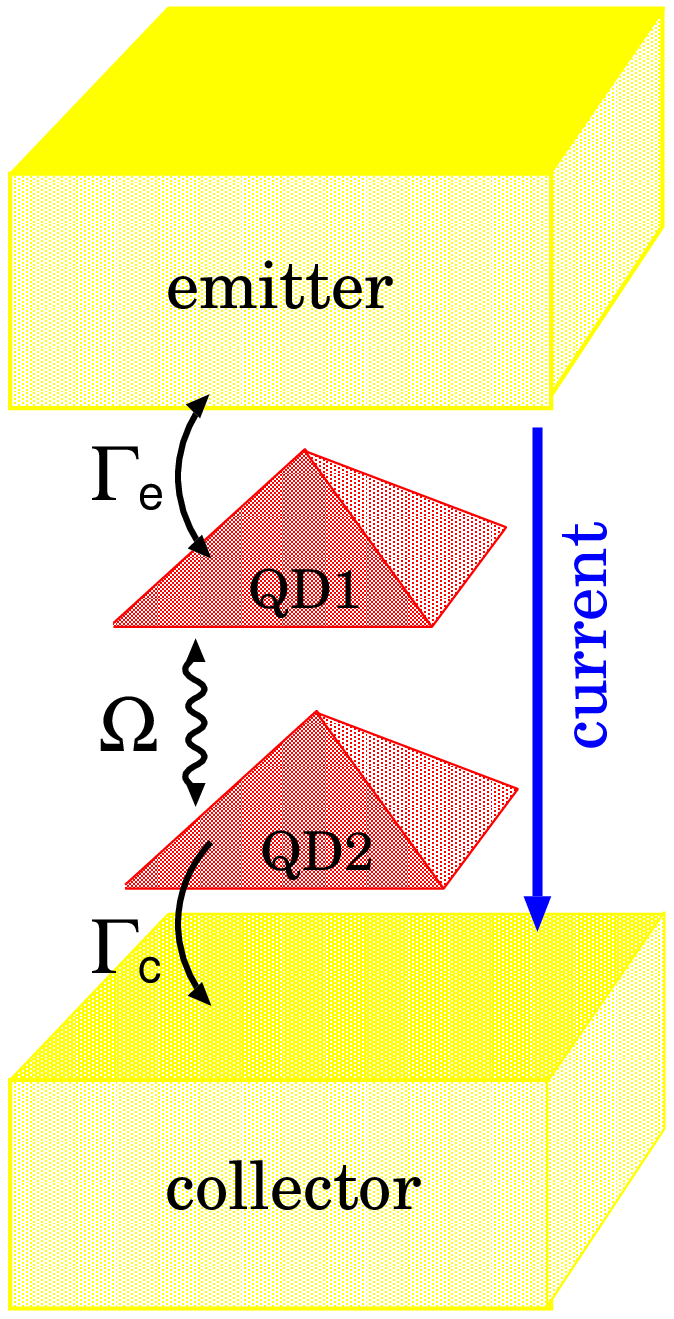}}
    \hspace{0.5cm}
    \includegraphics[width=0.18\textwidth]{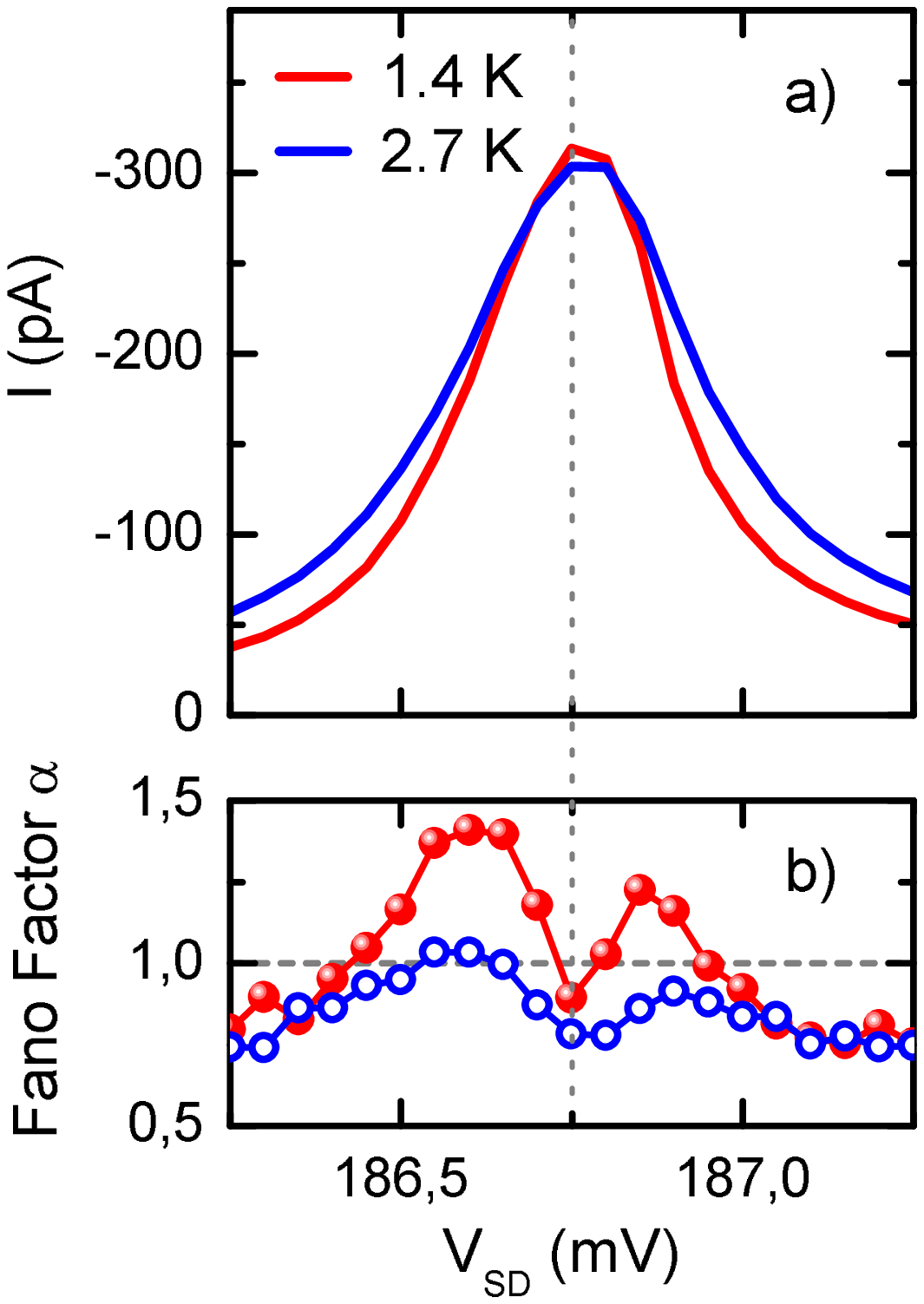}
    \caption{(Color online). Left: Sketch of a quantum dot
        stack with contact regions, contact couplings $\Gamma_e, \Gamma_c$ and the 
	tunnel matrix element $\Omega$ for coupling between the dots.
        The electrons tunnel from the
	top (emitter) reservoir via the dots into the bottom
        (collector) reservoir.
        Right: a) Measured current and b)
        Fano factor vs. bias voltage.}
	\label{fig1}
  \end{center}
\end{figure}

%%%%%%%%%%%%%%%%%%%%%%%%%%%%%%%%%%%%%%%%%%%%%%%%%%%%%%%%%%%%%%%%%%%%%%%%%%%%%%%%%%%%%%%%%%%%%%%%%%%%%%%

{\em Experiment.} Here we summarize the features of the device and
the experimental data \cite{BAR06} which are relevant for the
subsequent theoretical discussion. The InAs quantum dot stacks are
prepared by Stranski-Krastanov growth on AlAs and are sandwiched
between three AlAs tunneling barriers. The three-dimensional
electron reservoirs on both sides are formed in graded doped GaAs.
The pyramid-shaped quantum dots partially penetrate the middle
and top barrier which results in asymmetric tunneling rates.

A sketch of the device geometry is shown in Fig.~\ref{fig1}
(left). For the considered direction of the current -- from top to
bottom in Fig.~\ref{fig1} -- the emitter barrier is effectively
thinner, $\Gamma_e>\Gamma_c$. The leverage factors for the quantum
dots QD1 and QD2, i.e. the ratios of the potential drop between
emitter and dot to the applied voltage $V_{\rm SD}$, are estimated
from the device geometry: $\eta_1\approx 0.25$ and $\eta_2\approx 
0.55$, respectively. The Fermi energy in the reservoir can be
determined from a comparison with magneto-tunneling experiments
for devices with only one layer of quantum dots. For devices with
an otherwise identical growth scheme one finds an emitter Fermi
level $\mu_e\approx 13.6$meV~\cite{HAP00}.

Due to the small size of the InAs quantum dots on AlAs the ground
state energies $\varepsilon_1$ and $\varepsilon_2$ both lie above
the Fermi energy $\mu_e$ and no current can flow through the
device for $V_{\rm SD} = 0$. Due to the growth condition we also
find that the top quantum dot QD1 is larger than the bottom one,
$\varepsilon_1 < \varepsilon_2$. By applying a finite voltage
$V_{\rm SD} > 0$ the energy levels $\varepsilon_{i}(V_{\rm SD})$
of the dots are shifted downwards by $e\eta_{i}V_{\rm SD}$ and
brought into resonance. For $\mu_e > \varepsilon_{1} =
\varepsilon_{2} > 0$ we observe a peak in the tunneling current
(Fig.~\ref{fig1}a) \cite{VAA95}. 
With increasing temperature the resonance broadens while the maximum remains nearly constant. 
Also a slight asymmetry with respect to the maximum can be observed. 
Both facts provide some indication for electron-phonon scattering in the tunneling process \cite{FUJ98,BRA99} 
which will be discussed below in more detail.   

The noise properties of the current in the range of the resonant
tunneling peak is shown in Fig.~\ref{fig1}b. The Fano factor vs.
bias voltage displays an asymmetric double-peak structure with the maxima at
the edges of the current resonance and values larger than unity
indicating super-Poissonian noise. The noise is more sensitive to
the temperature than the current: for larger temperatures the Fano
factor peaks become smaller, and the super-Poissonian behavior
vanishes. Far from the resonance we observe a noise suppression below the
Poissonian value which we assume is due to some background current
signal originated from leackage channels. In an ideal device the
off-resonance electron transfer is Poissonian.    

%%%%%%%%%%%%%%%%%%%%%%%%%%%%%%%%%%%%%%%%%%%%%%%%%%%%%%%%%%%%%%%%%%%%%%%%%%%%%%%%%%%%%%%%%%%%%%%%%%%%%%%%%%%%

{\em Model.} We consider the coupled QD system as sketched in Fig.~\ref{fig1} (left) in the strong 
Coulomb blockade regime, i.e. only the occupation by a single excess electron is allowed. 
This restricts the dimension of the Hilbert space of the dot system spanned by the 
basis states $\vert 0\rangle\equiv\vert N,M\rangle$ for no excess electrons, 
$\vert 1\rangle\equiv\vert N+1,M\rangle$ for one excess electron in QD1 , 
and $\vert 2\rangle\equiv\vert N,M+1\rangle$ for one excess electron in QD2. 
The corresponding Hamiltonian can be written as

\begin{eqnarray}
H_D&=&\frac{\varepsilon}{2}\sigma_z +\Omega\sigma_x +\frac{1}{2}\sigma_z \hat{A}+H_B\nonumber\\
\hat{A}&=&\sum_Qg_Q(a_{-Q}+a_Q^\dagger ),\, H_B=\sum_Q\omega_Qa_Q^\dagger a_Q
\label{eq:hamilton}
\end{eqnarray}
with the detuning of the QD levels $\varepsilon\equiv\varepsilon_1-\varepsilon_2$ which corresponds 
to an applied bias voltage, the tunnel coupling between the dots $\Omega$ (see Fig.~\ref{fig1}), 
$\sigma_z=\vert 1\rangle\langle 1\vert -\vert 2\rangle\langle 2\vert$, and 
$\sigma_x=\vert 1\rangle\langle 2\vert +\vert 2\rangle\langle 1\vert$. 
The second part in (\ref{eq:hamilton}) describes the electron-phonon coupling and the phonon bath 
where $a_Q^\dagger$ is the creation operator of a phonon mode Q with frequency $\omega_Q$ and $g_Q$ 
is the coupling constant of electrons to phonon mode Q (for more details see Ref.~\cite{BRA02}). 
This spin-boson Hamiltonian will be coupled to the external electron reservoirs of emitter and collector contact.
The quantum master equation for the reduced density matrix 
is then obtained from the von-Neumann equation for the 
total density matrix  in Born-Markov approximation for weak contact coupling \cite{ELA02} and 
second-order perturbation theory in the electron-phonon coupling \cite{BRA02}:

\begin{eqnarray}
\frac{d}{dt}\rho =\left(
\begin{array}{ccccc}
-\Gamma_\textrm{e} & 0 & \Gamma_\textrm{c}e^{i\chi}  & 0 & 0\\
\Gamma_\textrm{e} & 0 & 0 & 0 & 2\Omega\\
0 & 0 & -\Gamma_c & 0 & -2\Omega \\
0 & \gamma_+ & -\gamma_- & -\frac{\Gamma_c}{2} -\gamma & -\varepsilon\\
0 & -\Omega & \Omega & \varepsilon & -\frac{\Gamma_c}{2} -\gamma
\end{array}
\right)\rho
\label{eq:master1}
\end{eqnarray}
with the reduced density matrix in vector form 
$\rho =\big(\rho_{00},\rho_{11},\rho_{22},\textrm{Re}[\rho_{12}],\textrm{Im}[\rho_{12}]\big)^T$ and the 
rates for coupling to the emitter/collector contact $\Gamma_{\textrm{e/c}}$, respectively (see Fig.~\ref{fig1} left).
The counting field $e^{i\chi}$ which will be needed to calculate the current and the noise 
(see below) enters the matrix element in (\ref{eq:master1}), where an electron jumps into the collector \cite{KIE05a}. 

The rates for the electron-phonon interaction in Eq.~(\ref{eq:master1}) are \cite{BRA02,BRA05}

\begin{eqnarray}
\gamma &=&\frac{g\pi}{\Delta^2}\left[\frac{\varepsilon^2}{\beta}+2\Omega^2\Delta e^{-\Delta /\omega_c}
\coth{\left(\frac{\beta\Delta}{2}\right)}\right]\nonumber\\
\gamma_{\pm} &=&g\frac{\pi\Omega}{\Delta^2}\left[\frac{\varepsilon}{\beta}
-\frac{\varepsilon}{2}\Delta e^{-\Delta /\omega_c}\coth{\left(\frac{\beta\Delta}{2}\right)
\mp\frac{\Delta^2}{2}e^{-\Delta /\omega_c}}\right]\nonumber\\
\end{eqnarray}
with a dimensionless coupling constant $g$, a high-frequency cutoff $\omega_c$ being an effective Debye frequency, 
$\Delta = \sqrt{\varepsilon^2+4\Omega^2}$ and $\beta = (k_BT)^{-1}$. 
These rates were derived for a bosonic environment with Ohmic spectral density 
$\rho (\omega )=g\omega e^{-\omega /\omega_c}\Theta (\omega )$ which corresponds to bulk piezoelectric 
phonons in the limit $\omega_c\rightarrow\infty$ and vanishing longitudinal speed of sound. 
Here and in the following we neglect the imaginary parts of $\gamma_\pm$.  

\begin{figure}[b]
  \begin{center}
    \includegraphics[width=0.267\textwidth]{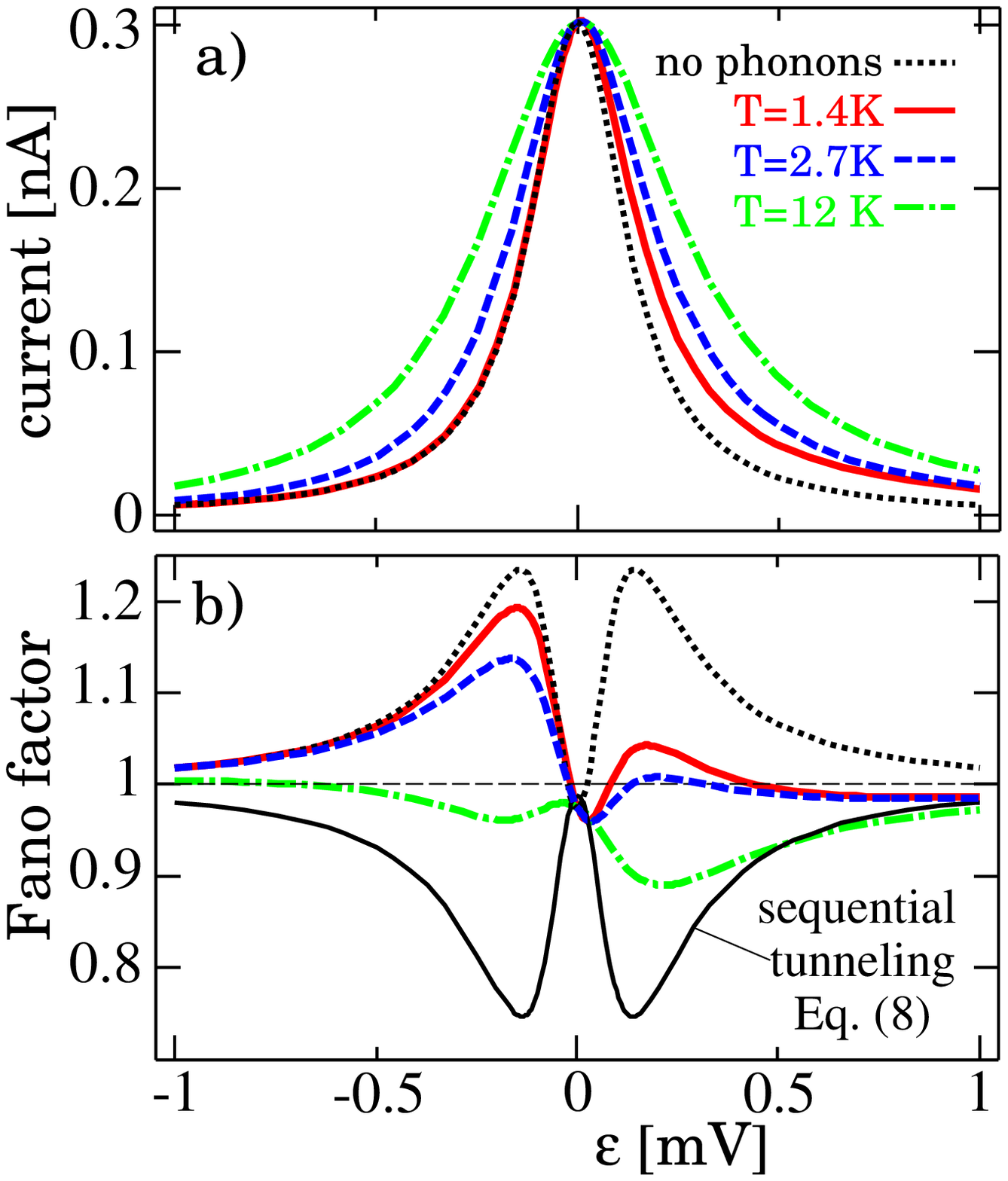}
    \includegraphics[width=0.21\textwidth]{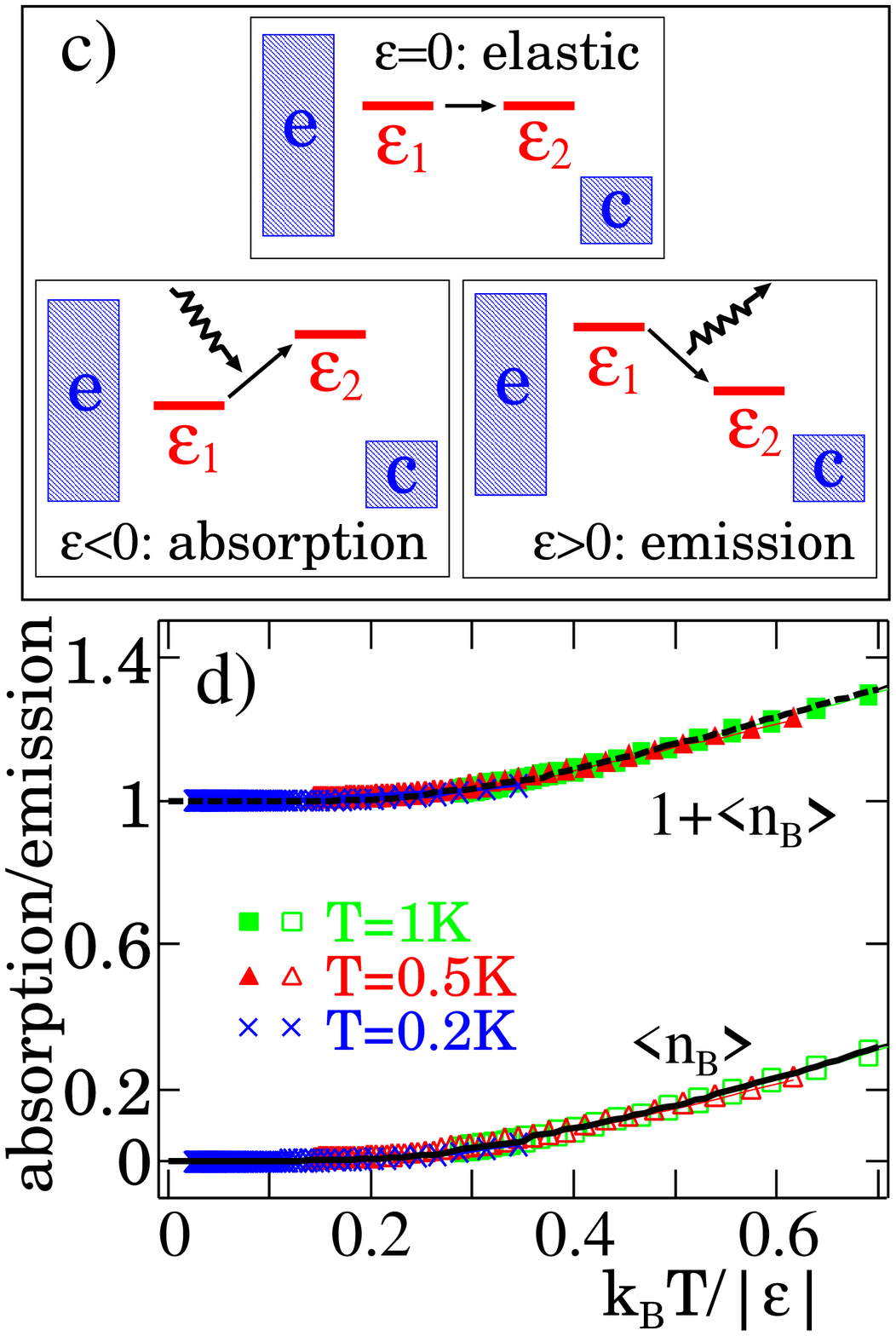}
    \caption{(Color online) Theory: a) Current, b) Fano factor vs. level detuning 
	$\varepsilon =\varepsilon_1-\varepsilon_2$ (bias voltage) for 
        strong Coulomb blockade (only one electron in the dot system). 
	c) Band schemes for tunneling through the dot system in different bias
        situations. d) absorption and emission rates normalized by the spontaneous emission rate 
	vs. $k_BT/\vert\varepsilon\vert$. The lower curve indicates the Bose-Einstein distribution
        $\langle n_B\rangle$, whereas the  upper curve shows $\langle n_B\rangle +1$. 
	Parameters: $\Gamma_e=0.1$meV, $\Gamma_c=2.5\mu$eV, $\Omega =0.1$meV, 
        $g=8\cdot 10^{-4}$, $\omega_c=5$meV.}
\label{fig2}
  \end{center}
\end{figure}

The average current and the noise is computed with help of the cumulant generating function $F(\chi )$. 
It is defined by 

 \begin{equation}
 \exp{\big[-F(\chi )\big]}=\sum_NP(N,t_0)\exp{\big[iN\chi\big]}
 \label{eq:cumulant}
 \end{equation}
with the distribution function $P(N,t_0)$ 
of the number $N$ of transferred charges in the time interval $t_0$.
We obtain  $F(\chi )$ as the eigenvalue of the transition matrix in Eq.~(\ref{eq:master1}) which approaches zero for 
$\chi =0$ \cite{KIE05a}. From the cumulant generating function the cumulants 
$C_k=-(-i\partial_\chi )^kF(\chi )\vert_{\chi =0}$ are obtained. 
Then, the average current is $\langle I\rangle =eC_1$, the zero-frequency noise is $S(0)=2e^2C_2$, 
and the Fano factor reads $C_2 /C_1$. It is smaller than unity
for sub-Poissonian and larger than unity for super-Poissonian charge transfer.

%%%%%%%%%%%%%%%%%%%%%%%%%%%%%%%%%%%%%%%%%%%%%%%%%%%%%%%%%%%%%%%%%%%%%%%%%%%%%%%%%%%%%%%%%%%%%%%%%%%%%%%

{\em Discussion.} Without electron-phonon interaction ($\gamma =\gamma_\pm =0$) we reproduce the known 
results for the current \cite{STO96,GUR96c}

\begin{equation}
\langle I\rangle = e\frac{4\Gamma_e\Gamma_c\Omega^2}{4\Omega^2(2\Gamma_e+\Gamma_c)
+\Gamma_e\Gamma_c^2+4\varepsilon^2\Gamma_e}
\label{eq:current}
\end{equation}
which provides a Lorentzian with respect to the detuning (bias voltage) $\varepsilon$ as shown 
by the dotted curve in Fig.~\ref{fig2}a and the Fano factor \cite{ELA02,BRA05}

\begin{equation}
\frac{S(0)}{2e\langle I\rangle} = 1-8\Gamma_e\Omega^2\frac{4\varepsilon^2(\Gamma_c-\Gamma_e)
+3\Gamma_e\Gamma_c^2+\Gamma_c^3+8\Gamma_c\Omega^2}
{\big[\Gamma_e\Gamma_c^2+4\Gamma_e\varepsilon^2+4\Omega^2(\Gamma_c+2\Gamma_e) \big]^2} 
\label{eq:fano}
\end{equation}
By close inspection of the Fano factor expression (\ref{eq:fano}), 
super-Poissonian charge transfer occurs when the second term becomes negative. 
This occurs for $\varepsilon\neq 0$ and $\Gamma_c<\Gamma_e$. In this case, the Coulomb interaction 
is more effective due to the smaller collector coupling, and two Fano factor peaks larger than unity 
and symmetric with respect to the current maximum appear (see dotted curve in Fig.~\ref{fig2}b). 
We emphasize that this super-Poissonian transport behavior is only obtained for (i) coherent coupling 
between the dots and for (ii) strong Coulomb blockade. In order to verify
this, we address these two issues in more detail.

(i) Consider a sequential tunneling approach, i.e. without non-diagonal elements of the density matrix: 
The master equation in strong Coulomb blockade for the diagonal elements of the density matrix 
$\tilde{\rho}=\big(\rho_{00},\rho_{11},\rho_{22}\big)^T$ reads

\begin{eqnarray}
\frac{d}{dt}\tilde{\rho}=
\left(
\begin{array}{ccc}
-\Gamma_\textrm{e} & 0 & \Gamma_\textrm{c}e^{i\chi}\\
\Gamma_\textrm{e} & -Z & Z\\
0 & Z & -(\Gamma_c+Z)
\end{array}
\right)\tilde{\rho}
\label{eq:master2}
\end{eqnarray}
where $Z\equiv\frac{4\Omega^2}{\Gamma_c}\big[1+(2\varepsilon /\Gamma_c)^2\big]^{-1}$
is 
Fermi's golden rule transition rate between states $\vert 1\rangle$ and $\vert 2\rangle$ \cite{SPR04}.
Note that due to the Coulomb blockade only the collector coupling enters this
rate. 
Current and noise are computed by means of the cumulant 
generating function (\ref{eq:cumulant}). Then this approach yields the same expression for the 
current (\ref{eq:current}) as the coherent description. The Fano factor is

\begin{equation}
\frac{S(0)}{2e\langle I\rangle} = 1-\frac{4\Omega^2\Gamma_e(2\Gamma_c+\Gamma_e)
(4\varepsilon^2+\Gamma_c^2)+32\Gamma_c\Gamma_e\Omega^4}
{\big[\Gamma_e\Gamma_c^2+4\Gamma_e\varepsilon^2+4\Omega^2(\Gamma_c+2\Gamma_e) \big]^2} 
\label{eq:fano-seq}
\end{equation}
Clearly, the sequential (incoherent) charge transfer turns out to be solely sub-Poissonian since 
the second term in (\ref{eq:fano-seq}) is always positive.

(ii) In order to relax the constraint of strong Coulomb blockade we extend the Hilbert space 
by the basis state $\vert 3\rangle\equiv\vert N+1,M+1\rangle$
for two excess electrons in the system. This leads to the master equation for the
density matrix 
$\hat{\rho} =\big(\rho_{00},\rho_{11},\rho_{22},\rho_{33},\textrm{Re}[\rho_{12}],\textrm{Im}[\rho_{12}]\big)^T$:

\begin{eqnarray}
\frac{d}{dt}\hat{\rho} =\left(
\begin{array}{cccccc}
-2\Gamma_\textrm{e} & 0 & 2\Gamma_\textrm{c}e^{i\chi}  & 0 & 0 & 0\\
\Gamma_\textrm{e} & 0 & 0 & 2\Gamma_\textrm{c}e^{i\chi} & 0 & 2\Omega\\
0 & 0 & -\Gamma & 0  & 0 & -2\Omega \\
0 & 0 & \Gamma_e & -\Gamma_c & 0 & 0\\
0 & 0 & 0 & 0 & -\frac{\Gamma}{2} & -\varepsilon\\
0 & -\Omega & \Omega & 0 & \varepsilon & -\frac{\Gamma}{2}
\end{array}
\right)\hat{\rho}
\label{eq:master3}
\end{eqnarray}
with $\Gamma =2\Gamma_e+\Gamma_c$. Here, for the sake of clarity  we neglect the
electron-phonon scattering. The resulting Fano factor also indicates overall sub-Poissonian 
transport behavior as we have checked numerically.
Furthermore we can exclude double occupation of one of the dots since it has a
rather low probability due to the large on site Coulomb energy in InAs
QDs ($U \sim 20$~meV).

We will now discuss the influence of the finite temperature: With $k_BT
\ll U$ 
we can assume $f_e(\varepsilon)=1$ and $f_c(\varepsilon)=0$
for the Fermi functions of the leads for a typical situation as sketched
in Fig. 2c. Therefore temperature acts solely due to the coupling to the
phonon bath.

With electron-phonon interaction the calculated current and Fano factor vs. level detuning $\varepsilon$ are 
shown in Fig.~\ref{fig2} a) and b) for three different temperatures. With increasing temperature the current 
resonance becomes asymmetrically broadened, whereas the maximum is not affected. 
For $\varepsilon <0$ phonon absorption and for $\varepsilon >0$  phonon emission take place, 
see Fig.~\ref{fig2}c. 
From the data for the computed current we extract the corresponding phonon 
absorption and emission rates normalized by the rate for spontaneous phonon emission 
$A(\varepsilon )$: 
$W_a(\varepsilon )/A(-\varepsilon )$ for $\varepsilon <0$ and  $W_e(\varepsilon )/A(\varepsilon )$ for 
$\varepsilon <0$, respectively. According to the Einstein relations they should equal the Bose-Einstein distribution  
$\langle n_B\rangle$ for absorption and $\langle n_B\rangle +1$ for emission. In Fig.~\ref{fig2}d 
we compare them as a function of $k_BT/\vert\varepsilon\vert$ \cite{FUJ98}. 

The double peaks in the Fano factor discussed above, which were symmetric without coupling to the 
phonon bath (dotted curve in Fig.~\ref{fig2}b), now become asymmetric with lower values above the resonance. 
For higher temperatures the Fano factor peaks decrease. This is in complete agreement with
our experimental data (Fig.~\ref{fig1}). This demonstrates that in the experiment, the transport of electrons 
through the dot system can be assumed to be partially coherent, strongly
indicating electron-phonon scattering as the dephasing mechanism.

Furthermore, the low-frequency noise behavior enables the clear identification of a sequential and a coherent tunneling regime.
In contrast to small temperatures, for higher $T=$12 K (dash-dotted curve in Fig.~\ref{fig2}b) the Fano factor 
indicates sub-Poissonian electron transfer for all bias voltages. Moreover, the Fano factor peaks at the
edges of the resonance in the coherent tunneling regime turn into minima. Such behavior is also found 
by our Fano factor expression for sequential tunneling (\ref{eq:fano-seq}).
A corresponding curve 
is shown in Fig.~\ref{fig2}b clearly revealing the double minimum feature. Hence a double peak in the Fano 
factor refers to a coherent tunneling regime, whereas a double minimum indicates a sequential tunneling regime.

\begin{figure}[t]
   \begin{center}
     \includegraphics[width=0.32\textwidth]{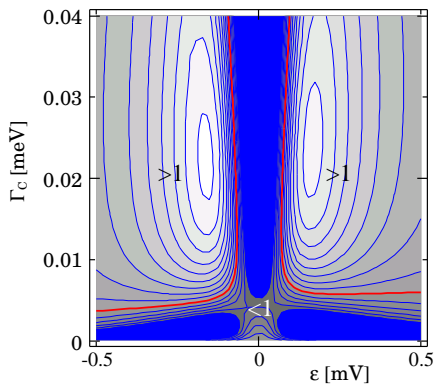}
     \caption{(Color online) Theory: Fano factor vs. level detuning $\varepsilon =\varepsilon_1-\varepsilon_2$ 
	(bias voltage) and collector coupling 
      $\Gamma_c$. Minimum/maximum Fano factor is 0.9/1.18, respectively. Contour line spacing is 0.02.
      Red curves correspond to unity Fano factor. $T=$12 K, 
	other parameters are the same as in Fig.~\ref{fig2}.}
     \label{fig3}
   \end{center}
 \end{figure}

The crossover between these two regimes or between sub- and
super-Poissonian charge transfer can also be induced
by varying the collector coupling $\Gamma_c$ for fixed couplings $\Gamma_e$,
$\Omega$ and temperature $T$ as shown in Fig.~\ref{fig3}.
The two different transport regimes are clearly distinguished: For small
coupling ($\Gamma_c <$ 5$\mu$eV) the sequential
tunneling regime is recognized by the double minima with suppressed
noise. For increasing $\Gamma_c$ a Fano factor
larger than unity for $\varepsilon\neq 0$ indicates positive
correlations in the charge transfer and consequently refers to
coherent tunneling. Here the increased $\Gamma_c$ leads to shorter dwell
time of the electrons on the dot and thus a smaller probability of a
phonon absorbtion as the bath temperature is kept constant. If we
increase $\Gamma_c$ above $\Gamma_e$ we find again a Fano factor smaller than unity.
Therefore the super-Poissonian behavior is most pronounced with a
maximum Fano factor for a certain collector coupling.

%%%%%%%%%%%%%%%%%%%%%%%%%%%%%%%%%%%%%%%%%%%%%%%%%%%%%%%%%%%%%%%%%%%%%%%%%%%%%%%%%%%%%%%%%%%%%%%%%%%%%%%

{\em Conclusions.} We have demonstrated that electron shot noise measurements provide a tool to detect the 
coherent coupling between quantum dots. 
We have compared low-frequency noise data with calculations based on quantum master equations which 
clearly demonstrate that the observed enhancement of the low-frequency noise at low temperatures is caused 
by the combined effect  of strong Coulomb blockade and the quantum coherent transfer of electrons between the dots. 
We have furthermore shown that electron-phonon scattering as a source of decoherence is responsible for the 
experimental temperature dependence of current and noise.

%%%%%%%%%%%%%%%%%%%%%%%%%%%%%%%%%%%%%%%%%%%%%%%%%%%%%%%%%%%%%%%%%%%%%%%%%%%%%%%%%%%%%%%%%%%%%%%%%%%%%%%%%

\acknowledgments We acknowledge helpful discussions with Alessandro Braggio. 
This work was supported by Deutsche Forschungsgemeinschaft in the framework of Sfb 296, 
project BR/1528/5-1/, the WE-Heraeus foundation and the BMBF, project nanoQuit.

%%%%%%%%%%%%%%%%%%%%%%%%%%%%%%%%%%%%%%%%%%%%%%%%%%%%%%%%%%%%%%%%%%%%%%%%%%%%%%%%%%%%%%%%%%%%%%%%%%%%%%%%


\vspace{-5mm}

\begin{thebibliography}{28}

\expandafter\ifx\csname natexlab\endcsname\relax\def\natexlab#1{#1}\fi
\expandafter\ifx\csname bibnamefont\endcsname\relax
  \def\bibnamefont#1{#1}\fi
\expandafter\ifx\csname bibfnamefont\endcsname\relax
  \def\bibfnamefont#1{#1}\fi
\expandafter\ifx\csname citenamefont\endcsname\relax
  \def\citenamefont#1{#1}\fi
\expandafter\ifx\csname url\endcsname\relax
  \def\url#1{\texttt{#1}}\fi
\expandafter\ifx\csname urlprefix\endcsname\relax\def\urlprefix{URL }\fi
\providecommand{\bibinfo}[2]{#2}
\providecommand{\eprint}[2][]{\url{#2}}

\bibitem[{\citenamefont{van~der Wiel et~al.}(2003)\citenamefont{van~der Wiel,
  Franceschi, Elzerman, Fujisawa, Tarucha, and Kouwenhoven}}]{WIE03}
\bibinfo{author}{\bibfnamefont{W.~G.} \bibnamefont{van~der Wiel}}
\bibinfo{author}{\bibfnamefont{{\em et al.}}},
  \bibinfo{journal}{Rev. Mod. Phys.} \textbf{\bibinfo{volume}{75}},
  \bibinfo{pages}{1} (\bibinfo{year}{2003}).


\bibitem[{\citenamefont{Hayashi et~al.}(2003)\citenamefont{Hayashi, Fujisawa,
  Cheong, Jeong, and Hirayama}}]{HAY03}
 \bibinfo{author}{\bibfnamefont{T.}~\bibnamefont{Hayashi}}
\bibinfo{author}{\bibfnamefont{{\em et al.}}},
  \bibinfo{journal}{Phys.~Rev.~Lett.} \textbf{\bibinfo{volume}{91}},
  \bibinfo{pages}{226804} (\bibinfo{year}{2003}).

\bibitem[{\citenamefont{Blanter and B{\"u}ttiker}(2000)}]{BLA00}
\bibinfo{author}{\bibfnamefont{Y.~M.} \bibnamefont{Blanter}} \bibnamefont{and}
  \bibinfo{author}{\bibfnamefont{M.}~\bibnamefont{B{\"u}ttiker}},
  \bibinfo{journal}{Phys.~Rep.} \textbf{\bibinfo{volume}{336}},
  \bibinfo{pages}{1} (\bibinfo{year}{2000}).

\bibitem[{\citenamefont{Aguado and Brandes}(2004)}]{AGU04}
\bibinfo{author}{\bibfnamefont{R.}~\bibnamefont{Aguado}} \bibnamefont{and}
  \bibinfo{author}{\bibfnamefont{T.}~\bibnamefont{Brandes}},
  \bibinfo{journal}{Phys.~Rev.~Lett.} \textbf{\bibinfo{volume}{92}},
  \bibinfo{pages}{206601} (\bibinfo{year}{2004}).

\bibitem[{\citenamefont{Fujisawa
  et~al.}(2006{\natexlab{a}})\citenamefont{Fujisawa, Hayashi, Tomita, and
  Hirayama}}]{FUJ06a}
\bibinfo{author}{\bibfnamefont{T.}~\bibnamefont{Fujisawa}},
\bibinfo{author}{\bibfnamefont{{\em et al.}}},
  \bibinfo{journal}{Science} \textbf{\bibinfo{volume}{312}},
  \bibinfo{pages}{1634} (\bibinfo{year}{2006}{\natexlab{a}}).



\bibitem[{\citenamefont{Barthold et~al.}(2006)\citenamefont{Barthold, Hohls,
  Maire, Pierz, and Haug}}]{BAR06}
\bibinfo{author}{\bibfnamefont{P.}~\bibnamefont{Barthold}}
\bibinfo{author}{\bibfnamefont{{\em et al.}}},
  \bibinfo{journal}{Phys.~Rev.~Lett.} \textbf{\bibinfo{volume}{96}},
  \bibinfo{pages}{246804} (\bibinfo{year}{2006}).

\bibitem[{\citenamefont{Iannaccone et~al.}(1998)\citenamefont{Iannaccone,
  Lombardi, Macucci, and Pellegrini}}]{IAN98}
\bibinfo{author}{\bibfnamefont{G.}~\bibnamefont{Iannaccone}}
\bibinfo{author}{\bibfnamefont{{\em et al.}}},
  \bibinfo{journal}{Phys.~Rev.~Lett.} \textbf{\bibinfo{volume}{80}},
  \bibinfo{pages}{1054} (\bibinfo{year}{1998}).

\bibitem[{\citenamefont{Kuznetsov et~al.}(1998)
\citenamefont{Kuznetsov, Mendez,M  Bruno, and Pham}}]{KUZ98}
\bibinfo{author}{\bibfnamefont{V.~V.} \bibnamefont{Kuznetsov}}
\bibinfo{author}{\bibfnamefont{{\em et al.}}},
 \bibinfo{journal}{Phys.~Rev.~B} \textbf{\bibinfo{volume}{58}},
  \bibinfo{pages}{R10159} (\bibinfo{year}{1998}).

\bibitem[{\citenamefont{Safonov et~al.}(2003)\citenamefont{Safonov, Savchenko,
  Bagrets, Jouravlev, Nazarov, Linfield, and Ritchie}}]{SAF03}
\bibinfo{author}{\bibfnamefont{S.~S.} \bibnamefont{Safonov}}
\bibinfo{author}{\bibfnamefont{{\em et al.}}},
   \bibinfo{journal}{Phys.~Rev.~Lett.}
  \textbf{\bibinfo{volume}{91}}, \bibinfo{pages}{136801}
  (\bibinfo{year}{2003}).

\bibitem[{\citenamefont{Gustavsson et~al.}(2006)\citenamefont{Gustavsson,
  Leturcq, Simovi{\u c}, Schleser, Ihn, Studerus, Ensslin, Driscoll, and
  Gossard}}]{GUS06a}
\bibinfo{author}{\bibfnamefont{S.}~\bibnamefont{Gustavsson}}
\bibinfo{author}{\bibfnamefont{{\em et al.}}},
   \bibinfo{journal}{Phys.~Rev.~B}
  \textbf{\bibinfo{volume}{74}}, \bibinfo{pages}{195305}
  (\bibinfo{year}{2006}).

\bibitem[{\citenamefont{Zarchin et~al.}(2007)\citenamefont{Zarchin, Chung,
  Heiblum, Rohrlich, and Umansky}}]{ZAR07}
\bibinfo{author}{\bibfnamefont{O.}~\bibnamefont{Zarchin}},
\bibinfo{author}{\bibfnamefont{{\em et al.}}},
  \bibinfo{journal}{Phys.~Rev.~Lett.} \textbf{\bibinfo{volume}{98}},
  \bibinfo{pages}{066801} (\bibinfo{year}{2007}).

\bibitem[{\citenamefont{Kie{\ss}lich et~al.}(2003)\citenamefont{Kie{\ss}lich,
  Wacker, and Sch{\"o}ll}}]{KIE03b}
\bibinfo{author}{\bibfnamefont{G.}~\bibnamefont{Kie{\ss}lich}},
  \bibinfo{author}{\bibfnamefont{A.}~\bibnamefont{Wacker}}, \bibnamefont{and}
  \bibinfo{author}{\bibfnamefont{E.}~\bibnamefont{Sch{\"o}ll}},
  \bibinfo{journal}{Phys.~Rev.~B} \textbf{\bibinfo{volume}{68}},
  \bibinfo{pages}{125320} (\bibinfo{year}{2003}).

\bibitem[{\citenamefont{Thielmann et~al.}(2005)\citenamefont{Thielmann,
  Hettler, K{\"o}nig, and Sch{\"o}n}}]{THI04}
\bibinfo{author}{\bibfnamefont{A.}~\bibnamefont{Thielmann}}
\bibinfo{author}{\bibfnamefont{{\em et al.}}},
  \bibinfo{journal}{Phys.~Rev.~B} \textbf{\bibinfo{volume}{71}},
  \bibinfo{pages}{045341} (\bibinfo{year}{2005}).

\bibitem[{\citenamefont{Cottet et~al.}(2004)\citenamefont{Cottet, Belzig, and
  Bruder}}]{COT04}
\bibinfo{author}{\bibfnamefont{A.}~\bibnamefont{Cottet}},
  \bibinfo{author}{\bibfnamefont{W.}~\bibnamefont{Belzig}}, \bibnamefont{and}
  \bibinfo{author}{\bibfnamefont{C.}~\bibnamefont{Bruder}},
  \bibinfo{journal}{Phys.~Rev.~B} \textbf{\bibinfo{volume}{70}},
  \bibinfo{pages}{115315} (\bibinfo{year}{2004}).

\bibitem[{\citenamefont{Belzig}(2005)}]{BEL05}
\bibinfo{author}{\bibfnamefont{W.}~\bibnamefont{Belzig}},
  \bibinfo{journal}{Phys.~Rev.~B} \textbf{\bibinfo{volume}{71}},
  \bibinfo{pages}{161301(R)} (\bibinfo{year}{2005}).

\bibitem[{\citenamefont{Djuric et~al.}(2005)\citenamefont{Djuric, Dong, and
  Cui}}]{DJU05}
\bibinfo{author}{\bibfnamefont{I.}~\bibnamefont{Djuric}},
  \bibinfo{author}{\bibfnamefont{B.}~\bibnamefont{Dong}}, \bibnamefont{and}
  \bibinfo{author}{\bibfnamefont{H.~L.} \bibnamefont{Cui}},
  \bibinfo{journal}{Appl.~Phys.~Lett.} \textbf{\bibinfo{volume}{87}},
  \bibinfo{pages}{032105} (\bibinfo{year}{2005}).

\bibitem[{\citenamefont{Aghassi
  et~al.}(2006{\natexlab{b}})\citenamefont{Aghassi, Thielmann, Hettler, and
  Sch{\"o}n}}]{AGH06a}
\bibinfo{author}{\bibfnamefont{J.}~\bibnamefont{Aghassi}},
\bibinfo{author}{\bibfnamefont{{\em et al.}}},
  \bibinfo{journal}{Phys.~Rev.~B} \textbf{\bibinfo{volume}{73}},
  \bibinfo{pages}{195323} (\bibinfo{year}{2006}{\natexlab{b}}).


\bibitem[{\citenamefont{Hapke-Wurst et~al.}(2000)\citenamefont{Hapke-Wurst,
  Zeitler, Frahm, Jansen, Haug, and Pierz}}]{HAP00}
\bibinfo{author}{\bibfnamefont{I.}~\bibnamefont{Hapke-Wurst}}
\bibinfo{author}{\bibfnamefont{{\em et al.}}},
  \bibinfo{journal}{Phys.~Rev.~B} \textbf{\bibinfo{volume}{62}},
  \bibinfo{pages}{12621} (\bibinfo{year}{2000}).

\bibitem[{\citenamefont{van~der Vaart et~al.}(1995)\citenamefont{van~der Vaart,
  Godijn, Nazarov, Harmans, Mooij, Molenkamp, and Foxon}}]{VAA95}
\bibinfo{author}{\bibfnamefont{N.}~\bibnamefont{van~der Vaart}}
\bibinfo{author}{\bibfnamefont{{\em et al.}}},
  \bibinfo{journal}{Phys.~Rev.~Lett.} \textbf{\bibinfo{volume}{74}},
  \bibinfo{pages}{4702} (\bibinfo{year}{1995}).

\bibitem[{\citenamefont{Fujisawa et~al.}(1998)\citenamefont{Fujisawa,
  Oosterkamp, van~der Wiel, Broer, Aguado, Tarucha, and Kouwenhoven}}]{FUJ98}
\bibinfo{author}{\bibfnamefont{T.}~\bibnamefont{Fujisawa}}
\bibinfo{author}{\bibfnamefont{{\em et al.}}},
  \bibinfo{journal}{Science} \textbf{\bibinfo{volume}{282}},
  \bibinfo{pages}{932} (\bibinfo{year}{1998}).

\bibitem[{\citenamefont{Brandes and Kramer}(1999)}]{BRA99}
\bibinfo{author}{\bibfnamefont{T.}~\bibnamefont{Brandes}} \bibnamefont{and}
  \bibinfo{author}{\bibfnamefont{B.}~\bibnamefont{Kramer}},
  \bibinfo{journal}{Phys.~Rev.~Lett.} \textbf{\bibinfo{volume}{83}},
  \bibinfo{pages}{3021} (\bibinfo{year}{1999}).

\bibitem[{\citenamefont{Brandes and Vorrath}(2002)}]{BRA02}
\bibinfo{author}{\bibfnamefont{T.}~\bibnamefont{Brandes}} \bibnamefont{and}
  \bibinfo{author}{\bibfnamefont{T.}~\bibnamefont{Vorrath}},
  \bibinfo{journal}{Phys.~Rev.~B} \textbf{\bibinfo{volume}{66}},
  \bibinfo{pages}{075341} (\bibinfo{year}{2002}).

\bibitem[{\citenamefont{Elattari and Gurvitz}(2002)}]{ELA02}
\bibinfo{author}{\bibfnamefont{B.}~\bibnamefont{Elattari}} \bibnamefont{and}
  \bibinfo{author}{\bibfnamefont{S.~A.} \bibnamefont{Gurvitz}},
  \bibinfo{journal}{Phys.~Lett.~A} \textbf{\bibinfo{volume}{292}},
  \bibinfo{pages}{289} (\bibinfo{year}{2002}).

\bibitem[{\citenamefont{Kie{\ss}lich et~al.}(2006)\citenamefont{Kie{\ss}lich,
  Samuelsson, Wacker, and Sch{\"o}ll}}]{KIE05a}
\bibinfo{author}{\bibfnamefont{G.}~\bibnamefont{Kie{\ss}lich}}
\bibinfo{author}{\bibfnamefont{{\em et al.}}},
  \bibinfo{journal}{Phys.~Rev.~B} \textbf{\bibinfo{volume}{73}},
  \bibinfo{pages}{033312} (\bibinfo{year}{2006}).

\bibitem[{\citenamefont{Brandes}(2005)}]{BRA05}
\bibinfo{author}{\bibfnamefont{T.}~\bibnamefont{Brandes}},
  \bibinfo{journal}{Phys.~Rep.} \textbf{\bibinfo{volume}{408}},
  \bibinfo{pages}{315} (\bibinfo{year}{2005}).

\bibitem[{\citenamefont{Stoof and Nazarov}(1996)}]{STO96}
\bibinfo{author}{\bibfnamefont{T.~H.} \bibnamefont{Stoof}} \bibnamefont{and}
  \bibinfo{author}{\bibfnamefont{Y.~V.} \bibnamefont{Nazarov}},
  \bibinfo{journal}{Phys.~Rev.~B} \textbf{\bibinfo{volume}{53}},
  \bibinfo{pages}{1050} (\bibinfo{year}{1996}).

\bibitem[{\citenamefont{Gurvitz and Prager}(1996)}]{GUR96c}
\bibinfo{author}{\bibfnamefont{S.~A.} \bibnamefont{Gurvitz}} \bibnamefont{and}
  \bibinfo{author}{\bibfnamefont{Y.~S.} \bibnamefont{Prager}},
  \bibinfo{journal}{Phys.~Rev.~B} \textbf{\bibinfo{volume}{53}},
 \bibinfo{pages}{15932} (\bibinfo{year}{1996}).

\bibitem[{\citenamefont{Sprekeler et~al.}(2004)\citenamefont{Sprekeler,
  Kie{\ss}lich, Wacker, and Sch{\"o}ll}}]{SPR04}
\bibinfo{author}{\bibfnamefont{H.}~\bibnamefont{Sprekeler}}
\bibinfo{author}{\bibfnamefont{{\em et al.}}},
  \bibinfo{journal}{Phys.~Rev.~B} \textbf{\bibinfo{volume}{69}},
  \bibinfo{pages}{125328} (\bibinfo{year}{2004}).

\end{thebibliography}
\end{document}